\documentclass[letterpaper,aps,prb,twocolumn,floatfix,amsfonts,amssymb,showpacs,superscriptaddress]{revtex4-1} 
\pdfpageattr {/Group << /S /Transparency /I true /CS /DeviceRGB>>}

\usepackage{amsmath} 
\usepackage{amsthm} 
\usepackage{amssymb} 
\usepackage{graphicx} 
\usepackage{tabularx} 
\usepackage{color} 
\usepackage{soul} 
\usepackage{subfigure}
\usepackage{sidecap}

\newcommand{\comment}[1]{}

\newcommand{\I}{\mathrm{i}} 
\newcommand{\E}{\mathrm{e}} 
\newcommand{\cre}[2]{#1_{#2}^\dagger} 
\newcommand{\ann}[2]{#1_{#2}^{\phantom{\dagger}}}

\newcommand{\boldk}{{\boldsymbol{\mathrm{k}}}} 
\newcommand{\boldq}{\boldsymbol{\mathrm{q}}} 
\newcommand{\boldr}{\boldsymbol{\mathrm{r}}}

\newcommand{\epsd}{\epsilon^{\phantom{\dagger}}_{d}}

\newcommand{\VBZ}{\Omega_{BZ}}

\newcommand{\sgn}[1]{\mathrm{sgn}(#1)}

\makeatletter 
\newsavebox{\@brx} 
\newcommand{\llangle}[1][]{\savebox{\@brx}{\(\m@th{#1\langle}\)}%
  \mathopen{\copy\@brx\kern-0.5\wd\@brx\usebox{\@brx}}} 
\newcommand{\rrangle}[1][]{\savebox{\@brx}{\(\m@th{#1\rangle}\)}%
  \mathclose{\copy\@brx\kern-0.5\wd\@brx\usebox{\@brx}}} 
\makeatother

\begin{document} 
\title{Multiple magnetic impurities on surfaces: scattering and quasiparticle interference}

\author{Andrew K. Mitchell} 
\affiliation{Oxford University, Department of Chemistry, Physical \& Theoretical Chemistry, South Parks Road, Oxford, OX1 3QZ, United Kingdom}
\affiliation{Institute for Theoretical Physics, Utrecht University, Leuvenlaan 4, 3584 CE Utrecht, The Netherlands}
\author{Philip G. Derry} 
\affiliation{Oxford University, Department of Chemistry, Physical \& Theoretical Chemistry, South Parks Road, Oxford, OX1 3QZ, United Kingdom}
\author{David E. Logan} 
\affiliation{Oxford University, Department of Chemistry, Physical \& Theoretical Chemistry, South Parks Road, Oxford, OX1 3QZ, United Kingdom} 
 

\begin{abstract} 
We study systems of multiple interacting quantum impurities deposited on a metallic surface in a 3d host. For the real-space two-impurity problem, using numerical renormalization group calculations, a rich range of behavior is shown to arise due to the interplay between Kondo physics and effective RKKY interactions --- provided the impurity separation is small. Such calculations allow identification of the minimum impurity separation required for a description in terms of independent impurities, and thereby the onset of the `dilute impurity limit' in many-impurity systems. A `dilute cluster' limit is also identified in systems with higher impurity density, where inter-impurity interactions are only important within independent clusters. We calculate the quasiparticle interference (QPI) due to two and many impurities, and explore the consequences of the independent impurity/cluster paradigms. Our results provide a framework to investigate the effects of disorder due to interacting impurities at experimentally relevant surface coverages.
\end{abstract}

\maketitle


\section{Introduction} 
\label{sec:intro} 

Electronic scattering due to impurities in metals gives rise to a variety of experimental signatures.\cite{hewson} Bulk measurements, such as resistivity, are of course affected; but \emph{local} quantities provide much more information about the effect of impurities. In particular, the development of scanning tunneling spectroscopy (STS) has allowed impurity positions/distributions on the surface to be imaged directly, and real-space maps of the surface local density of states (LDOS) to be built up.\cite{83:TersoffHamann,*87:FeenstraStroscio} These maps show pronounced inhomogeneities and even quantum interference due to the presence of impurities.\cite{CoonCu_Manoharan,*CoonCu_Nikolaus}

More recently, FT-STS -- the Fourier transform of LDOS maps obtained through STS -- has been connected to quasiparticle interference (QPI).\cite{98:PetersenHoffman,*00:PetersenHoffman,11:SimonBena} It describes how the Bloch state quasiparticles of a clean host system are scattered from impurities. Experimental QPI patterns are typically interpreted in terms of scattering from a single potential defect in the weak-scattering (Born) limit.\cite{11:SimonBena,03:CapriottiScalapino} The clean host structure can then be deduced from the preferred QPI scattering vectors. 

The phenomenon of impurity-induced scattering is much richer in the case of \emph{magnetic} impurities. Described in terms of a single interacting quantum level, such an impurity can host a free local moment at high energies/temperatures, which is then screened in metallic systems by conduction electrons through the Kondo effect\cite{hewson,wilson,*KWW} below a characteristic scale $T_K^{1\text{imp}}$. The enhanced spin-flip scattering responsible for the Kondo effect at low energies leads to a characteristic impurity spectral resonance\cite{hewson,02:GlossopLogan,asbasis,*fdmnrg} observable in STS.\cite{98:LiSchneiderWD,*09:TernesSchneiderWD,98:MadhavanCrommie,*01:MadhavanCrommie,wahl_dens} Induced real-space LDOS modulations from a single magnetic impurity produce a strong QPI signal, which acquires a universal temperature and scanning-energy dependence at low energies\cite{pd:qpi1} due to universality of the Kondo resonance in terms of $\omega/T_K^{1\text{imp}}$ and $T/T_K^{1\text{imp}}$. In the STS context, the scanning-energy $\omega=eV_{sd}$ is simply related to the bias voltage $V_{sd}$ between STM tip and surface.\cite{83:TersoffHamann,*87:FeenstraStroscio} The \emph{dynamics} of the QPI therefore provides information on the scattering t-matrix of the system.\cite{13:MitchellFritzTI,pd:qpi1} 

Unlike the case of a simple potential defect, interacting quantum impurities give rise to non-trivial many-body effects. The numerical renormalization group\cite{wilson,*KWW,nrg:rev} (NRG) is the theoretical method of choice for treating a single quantum impurity: it provides numerically-exact access to dynamical impurity quantities on essentially any temperature/energy scale.\cite{asbasis,*fdmnrg,asbasisprl} Solution of the local impurity problem then allows other real-space quantities\cite{11:MitchellBullaRS} and the QPI to be calculated, as detailed generally in Ref.~\onlinecite{pd:qpi1}.

However, real systems contain many impurities: the real-space surface region of a host material probed by STS to obtain QPI could include $N\sim 10-100$ impurities, depending on sampling size and impurity density. For potential defects, the physics remains rather trivial (the scattering t-matrix can be obtained in closed form\cite{pd:qpi1}). By contrast, systems containing many \emph{magnetic} impurities embody subtle interplays between Kondo physics, effective through-host RKKY interactions, and disorder in the impurity distribution. This is the scenario studied in the present work.

We specify a model in Sec.~\ref{sec:models} that takes into account explicitly the surface of a 3d metallic host onto which multiple magnetic impurities are deposited, allowing thereby calculation of surface LDOS maps and QPI. Realistic parameters are chosen, yielding single-impurity Kondo temperatures in the range $T_K^{1\text{imp}}\sim 0.1-100$K relevant to experiment. Electronic and quasiparticle scattering in this system is described in terms of the t-matrix, itself controlled by impurity dynamical quantities. The theoretical prescription for calculating QPI for a many-impurity system in terms of the impurity Green functions is given in Sec.~\ref{sec:dynamics}, including (exactly) the effect of surface quasiparticle dephasing by the host bulk.

Solution of such a model with $N$ interacting quantum impurities is a formidable challenge because in general it involves $N$ coupled screening channels. Generalization of NRG to deal with two spatially-separated impurities already represents a very significant increase in computational complexity, although accurate results for dynamical quantities are now possible. Our approach to the many-impurity problem in this paper begins with a detailed analysis of the two-impurity model, solved using NRG in Sec.~\ref{sec:2imp}. Inter-impurity interactions are important when the impurities are separated by a few lattice sites, giving rise to a rich range of correlated electron physics with distinctive QPI signatures.

A key finding, however, is that the impurities behave essentially \emph{independently} for surprisingly small inter-impurity separation. This has implications for the onset of the `dilute limit' in many-impurity systems. In this case, local quantities are in essence `blind' to the impurity distribution (although the impurity structure factor does affect the QPI). By analyzing random impurity distributions on the surface in Sec.~\ref{sec:QPImany}, we find that at experimentally-relevant impurity concentrations up to a few $\%$, there is a scale separation in the occurrence of impurity clusters of a given size $N_c$. This allows a description of many-impurity systems in terms of independent impurities and independent \emph{clusters}. As a representative example, we consider $N=100$ impurities randomly distributed in a $500\times 500$ surface sample region of a semi-infinite 3d cubic lattice host. On average, three $N_c=2$ clusters appear in such a sample: they are well-separated from other impurities or clusters and can be treated independently; and no clusters of size $N_c>2$ appear on average. Intra-cluster interactions are important for capturing local spectroscopic details; but the overall effect on QPI is weak due to low cluster occurrence at typical impurity densities.


\section{Model for multiple magnetic impurities} 
\label{sec:models} 

We consider $N>1$ magnetic impurities deposited on the surface of a metallic host system. The full Hamiltonian is given by
\begin{equation}
\label{eq:H}
H=H_{\text{host}}+ \sum^{N}_{\alpha=1} H_{\mathrm{imp},\alpha} \;.
\end{equation}
The host is taken here to be an infinite 3d cubic tight binding lattice, cleaved to reveal the (100) surface, 
\begin{equation}
\begin{split}
\label{eq:Hhost}
H_{\mathrm{host}} = -t\sum_{z=0}^{\infty} \Big [&\sum_{\langle ij\rangle,\sigma}(\cre{c}{\boldr_i z\sigma}\ann{c}{\boldr_j z \sigma} + \text{H.c.}) \\ + &\sum_{i,\sigma}(\cre{c}{\boldr_i z \sigma}\ann{c}{\boldr_i (z+1) \sigma} + \text{H.c.})  \Big ]  \;, 
\end{split}
\end{equation}
where $\cre{c}{\boldr_i z\sigma}$ creates an electron of spin $\sigma=\uparrow$/$\downarrow$ in the single Wannier orbital localized at site $\boldr_i$ of layer $z$, and $\langle ij\rangle$ denotes the sum over nearest-neighbor sites of a 2d square lattice layer. The resulting LDOS on the translationally-invariant surface is finite and flat at low energies, so although we focus on this specific host realization, the real-space physics is rather generic and typical of metallic 3d systems. More complicated materials can always be cast in the layered form of Eq.~\ref{eq:Hhost} by generalizing to a matrix structure.

We treat the surface explicitly to allow study of the QPI, which is necessarily obtained through surface measurement in FT-STS experiments. We note that coupling the surface layer to the semi-infinite bulk is required to capture the important effects of surface quasiparticle dephasing. Properties of the clean host are characterized by its real-space Green functions, here calculated exactly using the convolution method introduced in Ref.~\onlinecite{pd:qpi1}.

Each magnetic impurity $\alpha$ is described in terms of a single interacting quantum level, tunnel-coupled to a surface ($z=0$) host site at position $\boldr_{\alpha}$,
\begin{equation}
\label{eq:HAIM}
\begin{split}
H_{\mathrm{imp},\alpha} =& \sum_{\sigma}\epsd\cre{d}{\alpha\sigma}\ann{d}{\alpha\sigma} + U\cre{d}{\alpha\uparrow}\ann{d}{\alpha\uparrow}\cre{d}{\alpha\downarrow}\ann{d}{\alpha\downarrow} \\
+& V\sum_{\sigma}\left( \cre{d}{\alpha\sigma}\ann{c}{\boldr_\alpha 0 \sigma} + \text{H.c.} \right) \;, 
\end{split}
\end{equation}
where $\cre{d}{\alpha\sigma}$ creates a spin-$\sigma$ electron on impurity $\alpha$. Importantly, such an Anderson impurity has internal spin and charge dynamics due to the local electronic interactions. This is in contrast to a static 
`magnetic' impurity caricatured by an inhomogeneous magnetic field that breaks underlying time-reversal and spin SU(2) symmetry (and is non-interacting). The generalization of Eq.~\ref{eq:HAIM} to multi-orbital impurities is not considered in the present work  (the formal aspects of the scattering problem being largely unchanged).

The generalized quantum impurity problem involving $N$ magnetic impurities, spatially separated and coupled to conduction electrons of the host lattice, is distinctly nontrivial and exhibits a rich range of correlated electron physics. The simplest $N=1$ model, comprising a single magnetic impurity, already exhibits strong dynamical effects due to interactions, such as the Kondo effect.\cite{hewson,FeinAu,*FeinAuII} 

In the case of multiple impurities, richer physical behavior arises due to an effective RKKY interaction which couples impurities indirectly.\cite{54:RudermanKittel,*56:Kasuya,*57:Yosida,88:JonesVarmaWilkins,77:Doniach} The RKKY interaction is mediated via the host lattice and therefore depends on the specific impurity distribution in real-space. As shown explicitly in Sec.~\ref{sec:2imp_nrg}, impurities can be Kondo screened independently by the host, or entangled clusters can be collectively screened in a multi-stage process. Impurities can also `screen themselves' by forming inter-impurity singlet states when the effective RKKY interaction is strongly antiferromagnetic.


\subsection{Impurity parameters} 
\label{sec:params}

The Kondo physics of a \emph{single} magnetic impurity is sensitive to the underlying model parameters and host material through its LDOS.\cite{hewson,pd:qpi1} However, the RG flow and associated universality is controlled by an emergent energy scale $T_K^{1\text{imp}}$ --- the single-impurity Kondo temperature. In the Kondo limit,\cite{hewson} $T_K^{1\text{imp}}$ is determined by the interaction strength $U/\Gamma_0$, where $\Gamma_0=\pi V^2 \rho_0$ and $\rho_0=1/6t$ is the Fermi level surface LDOS of the 3d cubic lattice.

In real systems, Kondo temperatures measured in experiment are known to vary widely, even for a given impurity type and host material.\cite{impSTMrev} This is principally due to differences in the hybridization $\Gamma_0$, which is sensitive to details of the impurity's local environment. 

Guided by this, in the present work we use realistic, fixed values of the conduction electron bandwidth, $12t=11$~eV, and impurity interaction strength $U=3t=2.75$~eV (taken from Refs.~\onlinecite{solovyev,*00:UjsaghyZawadowski,*06:CastroNetoJones} for Co impurities on Au). For convenience, we consider $\epsilon_d=-U/2$ such that the impurity is singly-occupied. Appropriate choice of $V$ then gives rise to a realistic spread of single-impurity Kondo temperatures, as summarized in Table~\ref{tab:tk}.

\sidecaptionvpos{table}{}
\begin{SCtable}[2.5][h]
\begin{tabular}{| c | c |} 
\hline
$U/\Gamma_0$ & $T_K^{1\text{imp}}$ \\ [0.5ex]
\hline\hline 
12 & 85K  \\
16 & 15K  \\
20 & 3K   \\
24 & 0.5K \\
28 & 0.1K \\
\hline
\end{tabular}
\caption{Kondo temperatures for a single impurity on the 3d cubic lattice surface, with bandwidth $12t=11$~eV and $U=2.75$~eV. Calculated as the half-width at half-maximum of the Kondo spectral resonance at $T=0$ using NRG.}
\label{tab:tk}
\end{SCtable}

We note that $U/\Gamma_0$ in the range 12--16 yields Kondo temperatures consistent with classic studies of Co atoms on a Cu or Au surface;\cite{CoonCu_Manoharan,*CoonCu_Nikolaus,wahl_dens,impSTMrev,solovyev,*00:UjsaghyZawadowski,*06:CastroNetoJones,zitko_fano} while a passivating layer\cite{passivate} of $\text{Cu}_2$N between surface impurities and bulk reduces the hybridization to yield $T_K^{1\text{imp}}\approx 2$K. Indeed, $U/\Gamma_0=24$--$28$ is more appropriate for Fe in Au, where $T_K^{1\text{imp}}\approx 0.3$K.\cite{tk}

In the following we use these impurity parameters in the context of many-impurity systems, where different emergent energy scales and physics naturally arise.


\section{Electron and quasiparticle scattering} 
\label{sec:dynamics}

The impurity single-particle dynamics are described generically by elements of the Green function matrix $[\mathbf{G}_{d}(\omega)]_{\alpha,\beta}\equiv G^{\alpha\beta}_{d}(\omega) = \langle \langle \ann{d}{\alpha\sigma};\cre{d}{\beta\sigma}\rangle\rangle_{\omega}$, where $\langle \langle \hat{A};\hat{B}\rangle\rangle_{\omega}$ is the Fourier transform of the retarded correlator $-i\theta(t)\langle \{ \hat{A}(t),\hat{B}(0) \}\rangle$. The propagator $G^{\alpha\beta}_{d}(\omega)$ therefore contains information on the energy-dependent scattering of electrons between impurities $\alpha$ and $\beta$.

For a many-impurity system, such Green functions are obtained from a matrix Dyson equation,
\begin{equation}
\label{eq:dyson} 
\left[\mathbf{G}_{d}(\omega)\right]^{-1}=\left[\mathbf{g}_{d}(\omega)\right]^{-1}-\mathbf{\Sigma}(\omega) \;, 
\end{equation} 
where the non-interacting (but host-coupled) impurity Green functions are given by
\begin{equation} 
\label{eq:G0}
\left[\mathbf{g}_{d}(\omega)\right]^{-1}=(\omega+i0^{+}-\epsilon_d)\mathbf{I}-\mathbf{\Gamma}(\omega) \; ,
\end{equation} 
in terms of the hybridization matrix $\mathbf{\Gamma}(\omega)$ with elements $[\mathbf{\Gamma}(\omega)]_{\alpha,\beta} = V^2 G_{00}^0(\bold r_{\alpha},\bold r_{\beta},\omega) $. Here, $G_{zz'}^0(\bold r_{\alpha},\bold r_{\beta},\omega)=\langle \langle \ann{c}{\bold r_{\alpha}z\sigma};\cre{c}{\bold r_{\beta}z'\sigma}\rangle\rangle^0_{\omega}$ is the propagator between site $\bold r_{\alpha}$ of layer $z$ and $\bold r_{\beta}$ of layer $z'$ in the clean host (without impurities). $\bold r_{\alpha}$ and $\bold r_{\beta}$ are the host sites to which impurities $\alpha$ and $\beta$ are coupled on the surface layer, $z=0$. The self-energy matrix $\mathbf{\Sigma}(\omega)$ contains all information due to electronic interactions, which give rise to the Kondo effect, RKKY interaction, etc. 

Likewise, the full Green functions $G_{zz'}(\bold r_{i},\bold r_{j},\omega)$ describe electronic propagation through the host, in the presence of impurities. The contribution to the total scattering from the impurities is given by the real-space t-matrix equation,
\begin{eqnarray}
\label{eq:tmeqn_rs}
\begin{split}
G_{zz'}(\bold r_i,\bold r_j,\omega) =&G_{zz'}^0(\bold r_i,\bold r_j,\omega)   \\
+\sum_{\alpha,\beta} &G_{z0}^0(\bold r_i,\bold r_{\alpha},\omega) T_{\alpha\beta}(\omega) G_{0z'}^0(\bold r_{\beta},\bold r_{j},\omega) \;,
\end{split}
\end{eqnarray}
where the sum runs over \emph{impurity} sites $\alpha$ and $\beta$; and the t-matrix is given explicitly in terms of the impurity Green functions by
\begin{eqnarray}
\label{eq:tm_rs}
T_{\alpha\beta}(\omega)=V^2G_{d}^{\alpha\beta}(\omega) \;.
\end{eqnarray}

While host surface sites are of course coupled to those in the bulk, the STM tip itself probes only the host surface. In consequence, as detailed in Ref.~\onlinecite{pd:qpi1}, the scattering problem is most
effectively formulated in the so-called \emph{surface diagonal} basis, in which the layer index is preserved; leading to a description of the QPI in terms of \emph{surface} quasiparticles.
This follows from the \emph{partial} diagonalization of $H_{\mathrm{host}}$ (Eq.~\ref{eq:Hhost}) by 2d Fourier transformation of layers parallel to the surface,
\begin{equation}
\label{eq:partialdiag}
\ann{c}{\bold k_{\parallel} z \sigma} = \frac{1}{\VBZ^{1/2}} \sum_{\bold r_i} \E^{\I\boldr_i\cdot\bold{k_{\parallel}}}\ann{c}{\bold r_i z \sigma} \;,
\end{equation}
where $\VBZ$ is the volume of the first (surface) Brillouin zone (1BZ). The host Hamiltonian then reduces to a bundle of decoupled 1d chains,
\begin{equation}
\label{eq:Hhostkpar}
H_{\mathrm{host}} = \sum_{\bold k_{\parallel},\sigma} \Big [\sum_{z=0}^{\infty}\epsilon^{2d}_{\bold k_{\parallel}} \cre{c}{\bold k_{\parallel} z \sigma}\ann{c}{\bold k_{\parallel} z \sigma} -t(\cre{c}{\bold k_{\parallel} z \sigma}\ann{c}{\bold k_{\parallel} (z+1) \sigma} + \text{H.c.} ) \Big ]\;,
\end{equation}
where $\epsilon^{2d}_{\bold k_{\parallel}}=-2t[\cos(a_0 k_x)+\cos(a_0 k_y)]$ is the 2d square lattice dispersion, given in terms of surface momentum $\bold k_{\parallel}\equiv (k_x,k_y)$, and with lattice constant $a_0$. As a consequence, $G_z^{0}(\boldk_{\parallel},\boldk_{\parallel}',\omega)\equiv \langle \langle \ann{c}{\bold k_{\parallel} z \sigma};\cre{c}{\bold k_{\parallel}' z \sigma} \rangle\rangle_{\omega}^0 \propto \delta(\bold k_{\parallel}-\bold k_{\parallel}')$. Surface quasiparticles are however dephased by the bulk even in the clean host,\cite{pd:qpi1} with
\begin{equation}
\label{eq:G01d}
\begin{split}
G_{z=0}^{0}(\boldk_{\parallel},\omega) = f\left( \frac{\omega-\epsilon_{\bold k^{2d}_{\parallel}}}{2t}\right ) \qquad \text{where}\\
tf(\tilde{\omega}) = \tilde{\omega} -\begin{cases} \sgn{\tilde{\omega}}\sqrt{{\tilde{\omega}}^2-1}\quad &|\tilde{\omega}|>1\\\I\sqrt{1-\tilde{\omega}^2} \quad &|\tilde{\omega}|\leq1\end{cases} \;.
\end{split} 
\end{equation}
Scattering of surface quasiparticles due to impurities is then described\cite{pd:qpi1} by a surface t-matrix equation,
\begin{equation}
\begin{split}
\label{eq:tmeqn_kpar}
G_{0}(\boldk_{\parallel},\boldk'_{\parallel},\omega) =& G^0_{0}(\boldk_{\parallel},\omega)\delta(\bold k_{\parallel}-\bold k_{\parallel}') \\ &+  G_{0}^{0}(\boldk_{\parallel},\omega)T(\boldk_{\parallel},\boldk'_{\parallel},\omega)G_{0}^{0}(\boldk'_{\parallel},\omega) \;, 
\end{split}
\end{equation}
with the t-matrix itself given by,
\begin{equation} 
\label{eq:tm_kpar}
T(\boldk_{\parallel},\boldk'_{\parallel},\omega)= \frac{V^2}{\VBZ}\sum_{\alpha,\beta} \E^{\I(\bold k'\cdot \bold r_{\beta} - \bold k \cdot \bold r_{\alpha})} \times G_d^{\alpha\beta}(\omega)\;.
\end{equation}


\subsection{Quasiparticle interference} 
\label{sec:qpi}

Quasiparticle scattering from impurities gives rise to QPI, as measured in experiment by FT-STS.\cite{98:PetersenHoffman,*00:PetersenHoffman,*11:SimonBena} It is obtained from the LDOS map $\rho(\bold r_i, \omega)$ at scanning-energy $\omega$ and temperature $T$, measured in real-space over a sample region of size $L\times L$,
\begin{align}\label{eq:qpi_def}
\rho(\boldq,\omega)=\sum_{i\in(L \times L)}\E^{-\I\boldq\cdot\boldr_i}\rho(\boldr_i,\omega) \;.
\end{align}
 
For magnetic impurities deposited on the (100) surface of a semi-infinite 3d cubic lattice, the QPI can be calculated following Ref.~\onlinecite{pd:qpi1}. Taking the sample size $L\rightarrow \infty$ and subtracting the trivial contribution at $\bold q=\bold 0$ from the clean system, the exact QPI, $\Delta\rho(\boldq,\omega)=\rho(\boldq,\omega)-\rho^0(\boldq,\omega)$ from Eq.~\ref{eq:qpi_def} is given by,
\begin{eqnarray}
\label{eq:QPI_Q} 
\Delta\rho(\boldq,\omega)=-\frac{1}{2\pi \I} \left [ Q(\bold q,\omega) - Q(-\bold q, \omega)^* \right ] \;,
\end{eqnarray}
where 
\begin{eqnarray}
\label{eq:Q_def} 
\begin{split}
Q(\bold q,\omega) &= \int\limits_{1BZ}\frac{d^2\mathrm{\bold{k}_{\parallel}}}{\VBZ}~ \left [G_{0}(\boldk_{\parallel},\boldk_{\parallel}-\bold q,\omega) - G^0_{0}(\boldk_{\parallel},\omega)\delta(\bold q) \right ]\; \\
&\equiv V^2\sum_{\alpha,\beta}G_d^{\alpha\beta}(\omega)\times \Lambda_{\alpha\beta}(\bold q,\omega) \;
\end{split}
\end{eqnarray}
is given in terms of the full impurity Green functions $G_d^{\alpha\beta}(\omega)$. The quantity $\Lambda_{\alpha\beta}(\bold q,\omega)$ is defined for the \emph{clean} host material, and follows from Eqs.~\ref{eq:tmeqn_kpar} and \ref{eq:tm_kpar} as,
\begin{equation}
\label{eq:lambda_def} 
\begin{split}
\Lambda_{\alpha\beta}(\bold q,\omega) = \int\limits_{1BZ}\frac{d^2\mathrm{\bold{k}_{\parallel}}}{\VBZ}~&G_{0}^{0}(\boldk_{\parallel},\omega)G_{0}^{0}(\boldk_{\parallel}-\bold q,\omega) \\ &\times \E^{\I[\bold k_{\parallel} \cdot \bold r_{\alpha}- (\bold k_{\parallel}-\bold q) \cdot \bold r_{\beta}]} \;,
\end{split}
\end{equation}
which can be computed accurately and efficiently as a convolution.


\section{Two Impurities} 
\label{sec:2imp}

The two-impurity problem is the simplest to capture the competition between Kondo physics and inter-impurity interactions.\cite{81:JayaprakashKW,87:JonesVarma} A leading-order perturbative treatment generates the host-mediated RKKY interaction, coupling impurities by indirect exchange.\cite{54:RudermanKittel,*56:Kasuya,*57:Yosida,81:JayaprakashKW,spatiotemporalkondo} The magnitude and sign of the RKKY interaction depends sensitively on the dimensionality and geometry of the host,\cite{87:Abrahams} and on the impurity separation vector $\textbf{R}=\textbf{r}_2-\textbf{r}_1$.\cite{15:AllerdtBusserMartins}

Two spatially-separated Anderson impurities are often modelled by a two-impurity \emph{Kondo} model (2IKM),\cite{81:JayaprakashKW,87:JonesVarma} describing two exchange-coupled spin-$\tfrac{1}{2}$ impurities each coupled to its own independent conduction electron channel. The physics of this model is immensely rich: Kondo-screened and inter-impurity singlet phases arise, separated by a non-Fermi liquid quantum critical point\cite{89:JonesVarma,92:AffleckLudwig,95:AffleckLudwigJones,akm:2CKin2IK} of two-channel Kondo type.\cite{akm:2CKin2IK}

The 2IKM is however oversimplified, because the two conduction electron channels are not strictly independent in a true real-space system, both being constructed from states of the same electronic host.\cite{15:AllerdtBusserMartins} This is reflected by finite off-diagonal elements of the hybridization matrix, $\bold \Gamma(\omega)$ in Eq.~\ref{eq:G0}. Since $\Gamma_{12}(\omega)=V^2G_{00}^0(\bold r_{1},\bold r_{2},\omega)$ involves the host propagator between impurities located at sites $\bold r_{1} \ne \bold r_{2}$, the two-impurity problem evidently features inter-channel charge transfer processes. These processes are RG \emph{relevant}: simple perturbations that break the same symmetries can be incorporated directly into the 2IKM, and are known to destabilize the critical point.\cite{95:AffleckLudwigJones} Low-temperature/energy Fermi liquid crossovers then arise ubiquitously.\cite{akm:exactG,*akm:finiteT} 

The true real-space model does not therefore support a quantum phase transition. Instead, there is a crossover as a function of impurity separation or impurity-host coupling, with the Kondo regime evolving continuously into an RKKY-dominated regime.\cite{96:SilvaOliveiraWilkins,04:CampoOliveira,99:PaulaSilvaOliveira,11:ZhuZhu} This crossover is also known from the simpler two-impurity Anderson model, in which a hopping $t'$ directly tunnel-couples the impurities.\cite{92:Sakai1,*92:Sakai2,88:JonesVarmaWilkins,11:JayatilakaLogan} In that case, the RKKY interaction is generated to second-order in the tunnel-couplings, while the inter-channel charge transfer arises to third-order.\cite{11:JayatilakaLogan}

It is important to emphasize that inter-channel charge transfer processes cannot simply be neglected: finite $G_{00}^0(\bold r_{1},\bold r_{2},\omega)$ is the \emph{common origin} of both the RKKY interaction itself and the relevant perturbations destroying 2IKM criticality. Furthermore, the \emph{dynamical} nature of $G_{00}^0(\bold r_{1},\bold r_{2},\omega)$, resulting from \emph{through-lattice} electronic propagation, produces nontrivial RG flow. Physical regimes of the real-space model might therefore be inaccessible within the 2IAM (where $G_{00}^0(\bold r_{1},\bold r_{2},\omega)$ is replaced by a real constant hopping $t'$), or in the 2IKM (where it is neglected altogether).

As such, if one is interested in the real-space physics of two impurity systems --- and the resulting QPI --- one must study a real-space model directly. Here we solve the true real-space two-impurity problem exactly using 
NRG,\cite{wilson,*KWW,nrg:rev} as described below.


\subsection{Numerical Renormalization Group} 
\label{sec:nrg}

For two impurities separated in real space on the host surface, the first step is to diagonalize the matrix Dyson equation, Eq.~\ref{eq:dyson}. The translational invariance of the host surface implies that this can be achieved \emph{for all} $\omega$ by a single canonical transformation of operators to an even/odd orbital basis. This yields $G_{d}^{e/o}(\omega)=(\omega+\I 0^+ -\epsilon_d-\Gamma_{e/o}(\omega)  - \Sigma_{e/o})^{-1}$, in terms of even/odd quantities $\Omega_{e/o}=\Omega_{11}\pm \Omega_{12}$ (noting that $\Omega_{11}=\Omega_{22}$ and $\Omega_{12}=\Omega_{21}$). In the non-interacting system ($U=0$) where $\Sigma_{e/o}=0$, the even and odd channels are strictly decoupled: the even(odd) impurity combination couples only to the even(odd) host combination. However, the even/odd transformation scrambles the interaction term in $H_{\mathrm{imp}}$, coupling even and odd channels for $U\ne 0$, thereby requiring an irreducible \emph{two-channel} NRG calculation. The original information about the real-space separation of the impurities is encoded in the \emph{difference} between the dynamical quantities $\Gamma_{e}(\omega)$ and $\Gamma_{o}(\omega)$. As noted in Refs.~\onlinecite{95:AffleckLudwigJones}, \onlinecite{15:AllerdtBusserMartins}, the full energy dependence of these functions is required to capture the true real-space physics.

The \emph{explicit} transformation of real-space conduction electron operators to an even/odd basis (as described in Ref.~\onlinecite{spatiotemporalkondo}) can be highly complicated, depending on the host lattice and inter-impurity separation vector. We note however that this is not required, since the transformation can be performed on the level of the continuous, energy-dependent hybridization functions, as above.

The two-impurity NRG calculation then involves a logarithmic discretization\cite{wilson,*KWW,nrg:rev} of even/odd hybridization functions $\Gamma_{e/o}(\omega)$. A discretized version of the full model is then formulated in terms of even and odd impurity combinations, coupled to the end of even and odd Wilson chains. The model is then diagonalized iteratively, starting from a subsystem comprising the impurities themselves, and then building up the chains by successively coupling on even and odd Wilson chain orbitals. The couplings down each Wilson chain decrease exponentially due to the logarithmic discretization, and so high-energy states can be safely discarded at each step to avoid exponential Hilbert space growth. The physics of the problem is revealed on progressively lower energies/temperatures as the Wilson chains grow in length --- this is the essential RG character of the quantum impurity problem. 

The computational complexity of real-space two impurity problems is significantly greater\cite{akm:mint} than that of single impurity problems, which involve only a single effective conduction electron channel;
although accurate results are now possible. Further, impurity self-energies can be calculated accurately within NRG through the identity, 
\begin{equation}
\label{eq:SE}
\mathbf{\Sigma}(\omega) = [\textbf{G}_{d}(\omega)]^{-1}\textbf{F}_{d}(\omega) \;,
\end{equation}
where $[\textbf{F}_{d}(\omega)]_{\alpha,\beta}=U\langle\langle \ann{d}{\alpha\sigma};\cre{d}{\beta\sigma}\cre{d}{\beta\bar{\sigma}}\ann{d}{\beta\bar{\sigma}} \rangle\rangle_{\omega}$ (a matrix generalization to many impurities of the result
originally given for the single-impurity Anderson model\cite{UFG}).
Both $\textbf{G}_{d}(\omega)$ and $\textbf{F}_{d}(\omega)$ are calculated directly in NRG using the full density matrix approach\cite{asbasis,*fdmnrg} within the complete Anders-Schiller basis.\cite{asbasisprl} Impurity Green functions then follow from the matrix Dyson equation, Eq.~\ref{eq:dyson}.

In the following, the real-space two-impurity problem is solved with NRG using a discretization parameter $\Lambda=2$, and retaining $N_K=15,000$ states at each step of the iterative process. Conserved total charge and spin projection were exploited for block diagonalization. Convergence was established by increasing $\Lambda$ and $N_K$. $z$-averaging was not required.


\subsection{Effect of impurity separation} 
\label{sec:2imp_nrg}

\sidecaptionvpos{figure}{}
\begin{SCfigure*}
\includegraphics[width=13cm]{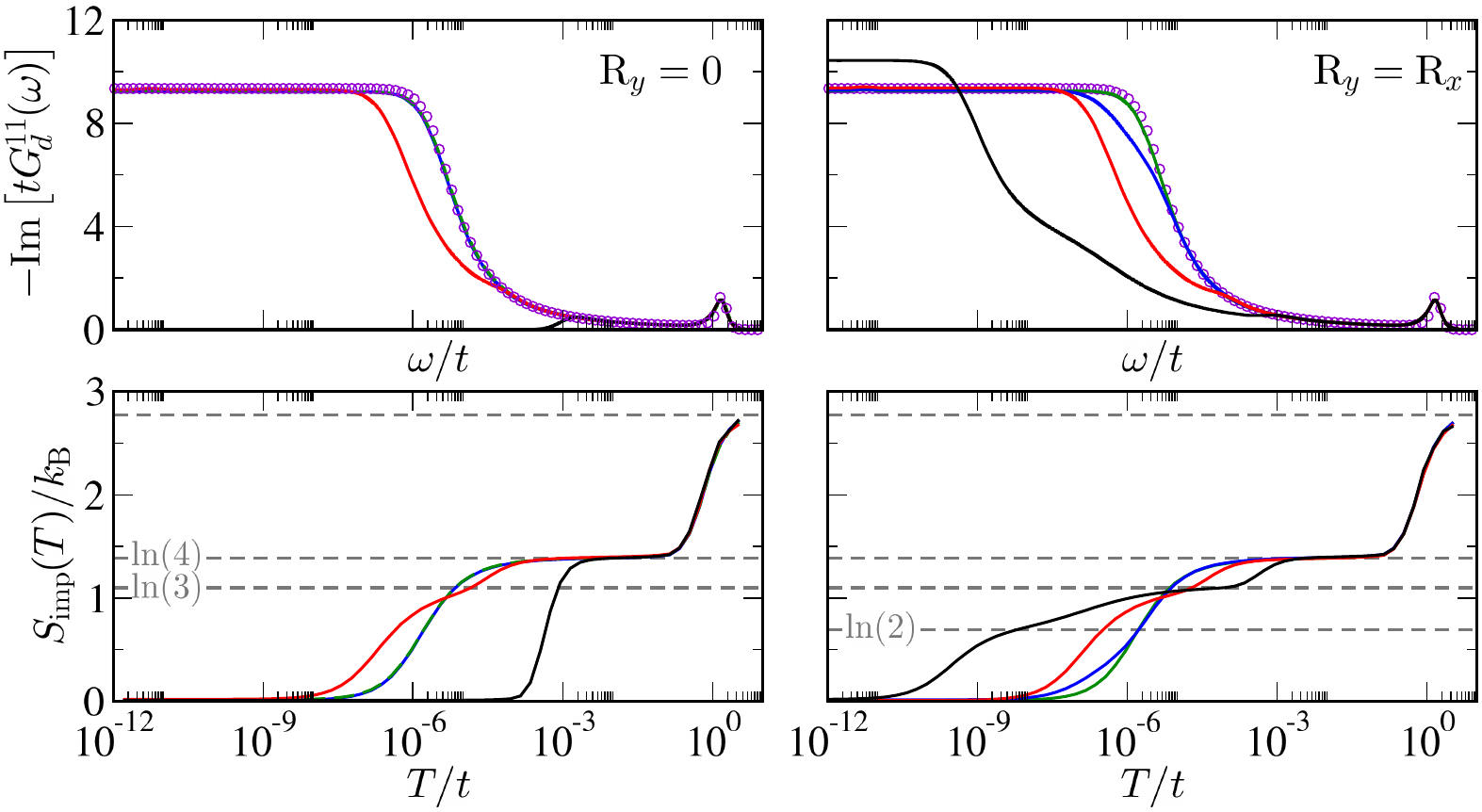}
\caption{\label{fig:2imp}
Two equivalent magnetic impurities on the 3d cubic (100) surface with $U/\Gamma_0=28$ (see Table~\ref{tab:tk}). NRG results for the local impurity spectrum $-\text{Im}[tG_{d}^{11}(\omega)]$ vs $\omega/t$ at $T=0$ (upper panels) and the temperature-dependence of the impurity contribution to entropy $S_{\text{imp}}(T)$ vs $T/t$ for the same systems (lower panels). Inter-impurity vector $\textbf{R}$, with $\text{R}_y=0$ (left panels) and $\text{R}_y=\text{R}_x$ (right), where $\text{R}_x/\text{a}_0=1,2,3,4$ for black, red, blue and green lines.  Circle points are the single-impurity result.}
\end{SCfigure*}

The interplay between Kondo physics and the effective host-mediated interaction between two magnetic impurities on a surface, gives rise to a rich range of physics. Aside from the bare energy scales $t$, $U$ and $V$ entering the problem, two \emph{emergent} scales arise: the single-impurity Kondo temperature $T_K^{1\text{imp}}$ and the effective inter-impurity RKKY exchange coupling, $J_{\text{RKKY}}$. The relative size of these scales, and the sign of $J_{\text{RKKY}}$, controls the underlying RG flow and low-energy physics.\cite{81:JayaprakashKW,87:JonesVarma} In the following, we use NRG to solve the two impurity problem, allowing these scales to be extracted for given impurity parameters and inter-impurity separation vector $\textbf{R}$. However, it is also instructive to obtain simple perturbative estimates. For a single impurity on the cubic lattice surface, third-order perturbative scaling\cite{anderson_pms,hewson} at particle-hole symmetry yields,
\begin{equation}
\label{eq:tk}
T_K^{1\text{imp}} = a t~\sqrt{\frac{\Gamma_0}{U}}\exp\left(-\frac{\pi}{8}\times\frac{U}{\Gamma_0}\right) \;,
\end{equation}
where $\Gamma_0=\pi V^2/6t$, as in Sec.~\ref{sec:params}. The estimates compare well with exact (non-perturbative) results in Table~\ref{tab:tk} using $a\approx 3$. For two impurities, the RKKY coupling is given to second order in $\Gamma_0/U$ as,\cite{spatiotemporalkondo}
\begin{eqnarray}
\label{eq:rkky}
\begin{split}
J_{\text{RKKY}} =& \left ( \frac{48 t}{\pi^2} \times\frac{\Gamma_0}{U} \right )^2 \\ \times & P\int_{-6t}^{0}d\epsilon \int_{0}^{6t} d \epsilon'~ \frac{\text{Im}G_{00}^0(\bold R,\epsilon) \text{Im}G_{00}^0(\bold R,\epsilon')}{\epsilon-\epsilon'} \;,
\end{split}
\end{eqnarray}
where $P$ denotes the principal value, and $G_{00}^0(\bold R,\epsilon)\equiv G_{00}^0(\bold r_1,\bold r_2,\epsilon)$ are free host Green functions connecting surface impurity sites $\bold r_1$ and $\bold r_2$. At large inter-impurity lattice separations, we find from Eq.~\ref{eq:rkky} that the RKKY coupling decays rapidly, faster than $1/|\bold R|^3$.

As such, when the impurities are well-separated, one expects $T_K^{1\text{imp}}\gg J_{\text{RKKY}}$ for a given $U/\Gamma_0$. Each impurity is then individually Kondo-screened essentially independently of the other, thereby quenching their magnetic moments and rendering the RKKY exchange interaction inoperative. By contrast, at small inter-impurity separations such that $J_{\text{RKKY}} \gg T_K^{1\text{imp}}$, RKKY physics should dominate, causing the impurities to bind together in either singlet or triplet configurations (depending on whether $J_{\text{RKKY}}$ is positive or negative). In this scenario, the single-impurity RG flow is cut off on the scale of $J_{\text{RKKY}}$, and single-impurity Kondo physics is destroyed. We remark however that, even in the antiferromagnetic case, the inter-impurity singlet remains entangled with the host, which mediates the indirect exchange coupling (in contrast to the situation in the 2IKM with direct exchange, where the singlet decouples).\cite{pd:2imp}

\sidecaptionvpos{table}{}
\begin{SCtable}[2.5][h]
\begin{tabular}{| c | c | c |} 
\hline
$R_x$ & $R_y=0$ & $R_y=R_x$ \\ [0.5ex]
\hline\hline 
1 & $+4.5\times 10^{-2}$ & $-2.5\times 10^{-2}$ \\
2 & $-2.5\times 10^{-3}$ & $-2.0\times 10^{-3}$ \\
3 & $+1.0\times 10^{-4}$ & $-2.9\times 10^{-5}$ \\
4 & $-2.5\times 10^{-6}$ & $-3.2\times 10^{-5}$ \\
\hline
\end{tabular}
\caption{Perturbative estimate of the RKKY coupling from Eq.~\ref{eq:rkky}, given as $(J_{\text{RKKY}}/t)\times (U/\Gamma_0)^2$, for inter-impurity separations $\bold R=(R_x,R_y)$ as per Fig.~\ref{fig:2imp}.
}
\label{tab:J}
\end{SCtable}

Table~\ref{tab:J} shows the perturbative estimate for the RKKY coupling, as obtained from numerical evaluation of Eq.~\ref{eq:rkky}, for various inter-impurity separations. For separations with $R_y=0$, there is a clear alternation between antiferromagnetic and ferromagnetic coupling; while for $R_y=R_x$, only ferromagnetic coupling arises. The rapid decay of $J_{\text{RKKY}}$ with increasing separation $|\bold R|$ is also clearly seen (even for small separations) from this simple calculation. 
The true variation of $J_{\text{RKKY}}$ --- and its subtle interplay with Kondo physics --- must however be extracted from the full NRG solution of the two-impurity model, which takes into account the dynamical nature of the coupling and renormalization effects. We now turn to NRG to examine these issues in detail.

Physical quantities, such as impurity dynamics and thermodynamics, provide detailed information about the underlying physics and screening mechanisms. In Fig.~\ref{fig:2imp} we use NRG to study the impurity spectrum 
$-\text{Im}[tG_d^{11}(\omega)]$, plotted vs $\omega/t$ at temperature $T=0$ in the upper panels; and the impurity contribution to entropy $S_{\text{imp}}(T)/k_\text{B}$, shown vs temperature $T/t$ in the lower panels. The two impurities are deposited on the cubic lattice surface, separated by a vector $\bold R\equiv (\text{R}_x,\text{R}_y)$; with $\text{R}_y=0$ considered in the left panels and $\text{R}_y=\text{R}_x$ in the right panels, and with $\text{R}_x/a_0=1,2,3,4$ in each case. Note that $G_d^{11}(\omega)=G_d^{22}(\omega)$ since the impurities are equivalent. The local spectra are related to the local parts of the scattering t-matrix through Eq.~\ref{eq:tm_rs}. 

We consider first the spectra in the upper panels. In all cases, a Hubbard satellite peak arises at high energies $|\omega|\sim U/2$, corresponding to impurity charge fluctuations.\cite{hewson} For $|\omega|\ll U$, the impurities become essentially singly-occupied and magnetic moments form. Although the impurities are coupled by the effective RKKY interaction, at energies or temperatures $\gg J_{\text{RKKY}}$, the impurities behave essentially independently. Corrections to the local moment fixed point then control the physics; incipient RG flow for each impurity towards a Kondo-screened strong-coupling state produces the classic inverse logarithmic spectral behavior\cite{02:GlossopLogan} when $\max(|J_{\text{RKKY}}|,T^{1\text{imp}}_K) \ll |\omega|\ll U$,
\begin{eqnarray}
\label{eq:LM}
-\text{Im}[tG_d^{11}(\omega)] \sim \frac{1}{1+b\ln^2|c~\omega/T^{1\text{imp}}_K|} \;,
\end{eqnarray}
with $b,c$ constants of $\mathcal{O}(1)$.

Consider now the $\text{R}_y =0$ cases (left panels, Fig.~\ref{fig:2imp}). For neighboring impurities ($\text{R}_x/\mathrm{a}_0=1$, black lines) the RKKY interaction $J_{\text{RKKY}}>0$ is antiferromagnetic and comparatively strong, causing the impurities to lock up together into a spin singlet state on the scale of $J_{\text{RKKY}}$. The RG flow toward independent Kondo strong coupling states is arrested on this scale since all impurity degrees of freedom are quenched. This results in an impurity spectrum $\text{Im}[tG_{d}^{11}]\simeq 0$ for $|\omega|\ll J_{\text{RKKY}}$. Confirming this physical picture, the impurity contribution to entropy shown in the lower panel is quenched to $S_{\text{imp}}=0$ on the same scale (for $T\gg |J_{\text{RKKY}}|$, $S_{\text{imp}}\sim k_\text{B}\ln(4)$, corresponding to two free impurity local moments). The effective RKKY coupling in this system can thus be extracted from these crossovers as $J_{\text{RKKY}}\approx 10^{-3}t$.

By contrast, when $\text{R}_x/\mathrm{a}_0=2$ (left panels, red lines), the effective RKKY interaction is ferromagnetic and weaker. On the scale of $|J_{\text{RKKY}}|\approx 10^{-4}t$, the impurities form a triplet configuration with an associated $\ln(3)$ impurity entropy. On lower temperature/energy scales $T_K^{S=1}\approx 10^{-6}t$, we find that the inter-impurity triplet is exactly Kondo screened~\cite{Nozieres1980} by even and odd conduction electron channels. On energy scales $T_K^{S=1}\ll |\omega| \ll |J_{\text{RKKY}}|$, spin-1 local moment physics naturally arises, with spectral behavior again that of Eq.~\ref{eq:LM}, but with the characteristic scale $T^{1\text{imp}}_K$ replaced by $T_K^{S=1}$. In fact, universal lineshapes are obtained as a function of $|\omega|/T_K^{S=1}$ and $T/T_K^{S=1}$ for all $|\omega|$ and $T\ll |J_{\text{RKKY}}|$. In particular, the lowest-energy behavior is characteristic of flow to a Fermi liquid ground state, with a fully quenched entropy, $S_{\text{imp}}(0)=0$. The low-energy behavior of the impurity spectrum has the characteristic Fermi liquid asymptotics,
\begin{equation}
\label{eq:FLspec}
-\text{Im}[tG_d^{11}(\omega)] ~\overset{|\omega| \ll T_K}{\sim}~ p-q\left(\frac{\omega}{T_K}\right )^2 \;,
\end{equation}
where $T_K\equiv T_K^{S=1}$ is the triplet Kondo scale, $q=\mathcal{O}(1)$ is a constant, and the Fermi level value $p$ depends on the full self-energy of the two-impurity problem (there is no simple pinning condition, as for the single-impurity case).\cite{pd:2imp}

On further separating the impurities, the RKKY interaction becomes even weaker. In fact, $|J_{\text{RKKY}}| \ll T_K^{1\text{imp}} \approx 10^{-5}t$ already for $\text{R}_x/\mathrm{a}_0=3$, $4$ -- left panels, blue and green lines (which are virtually indistinguishable) -- such that each impurity is \emph{independently} Kondo screened below the single-channel scale $T_K^{1\text{imp}}$. The lineshapes are again universal,\cite{hewson} although different from the triplet case in detail. Comparison with the circle points shows that the physics of each impurity is indeed that of a single impurity in the same host. Here the impurity spectrum is also described by Eq.~\ref{eq:FLspec}, but now with $p=\Gamma_0/t$ pinned by the Friedel sum rule.\cite{hewson,lutt}

Along $\text{R}_y=\text{R}_x$ (Fig.~\ref{fig:2imp}, right panels), one might expect the effective RKKY coupling to be ferromagnetic, since the impurities are always connected by an even number of hops on the lattice. As demonstrated by Fig.~\ref{fig:2imp}, this is indeed the case for $\text{R}_x/\mathrm{a}_0=1$ and $2$. An inter-impurity triplet thus forms in both cases, with incipient RG flow near $S_{\text{imp}}(T)=k_\text{B}\ln(3)$. For $\text{R}_x/\mathrm{a}_0=1$ (black lines), the Kondo screening of the inter-impurity triplet is distinctly \emph{two-stage}, indicating a strong inequivalence of the even and odd conduction electron channels. This results from significant off-diagonal elements of the hybridization matrix $\bold \Gamma(\omega)$ in Eq.~\ref{eq:G0} when the impurities are proximal. However, this anisotropy is already very small when $\text{R}_x/\mathrm{a}_0=2$ (red lines), so that the two-channel Kondo screening of the inter-impurity triplet proceeds essentially in a single step. At greater separations $\text{R}_x/\mathrm{a}_0=3$ and $4$ (blue and green lines), the magnitude of the RKKY interaction is again small enough that the two impurities remain essentially independent, and are separately Kondo screened.\\

For two impurities on the surface of a 3d cubic lattice, the overall results of Fig.~\ref{fig:2imp} shows that the dynamics are strongly affected when their separation is on the order of a few lattice sites. Importantly, however, the impurities are found to behave essentially independently for rather small separations, $|\bold R |\gtrsim 4 \sqrt{2}\mathrm{a}_0$. This is in accord with recent theoretical work,\cite{15:AllerdtBusserMartins} as well as experimental observation,\cite{CoonCu_Manoharan,*CoonCu_Nikolaus,impSTMrev,passivate,wahl_dens} where systems with small average impurity separation appear to be described in terms of the independent impurity paradigm. From a real-space perspective this might seem puzzling. A so-called `Kondo cloud' of conduction electrons surrounds each impurity and is responsible for screening its magnetic moment. This screening cloud is typically large, characterized\cite{08:AffleckSaleur,11:MitchellBullaRS,10:BusserMartins} by a length scale $\xi_K^{1\text{imp}} \sim \hbar v_F/k_B T_K^{1\text{imp}}$ (with $v_F$ the Fermi velocity). Our results show that such Kondo clouds can be substantially interpenetrating, even though the screening itself remains independent.\cite{note:clouds}


\sidecaptionvpos{figure}{}
\begin{figure}
\includegraphics[width=8.5cm]{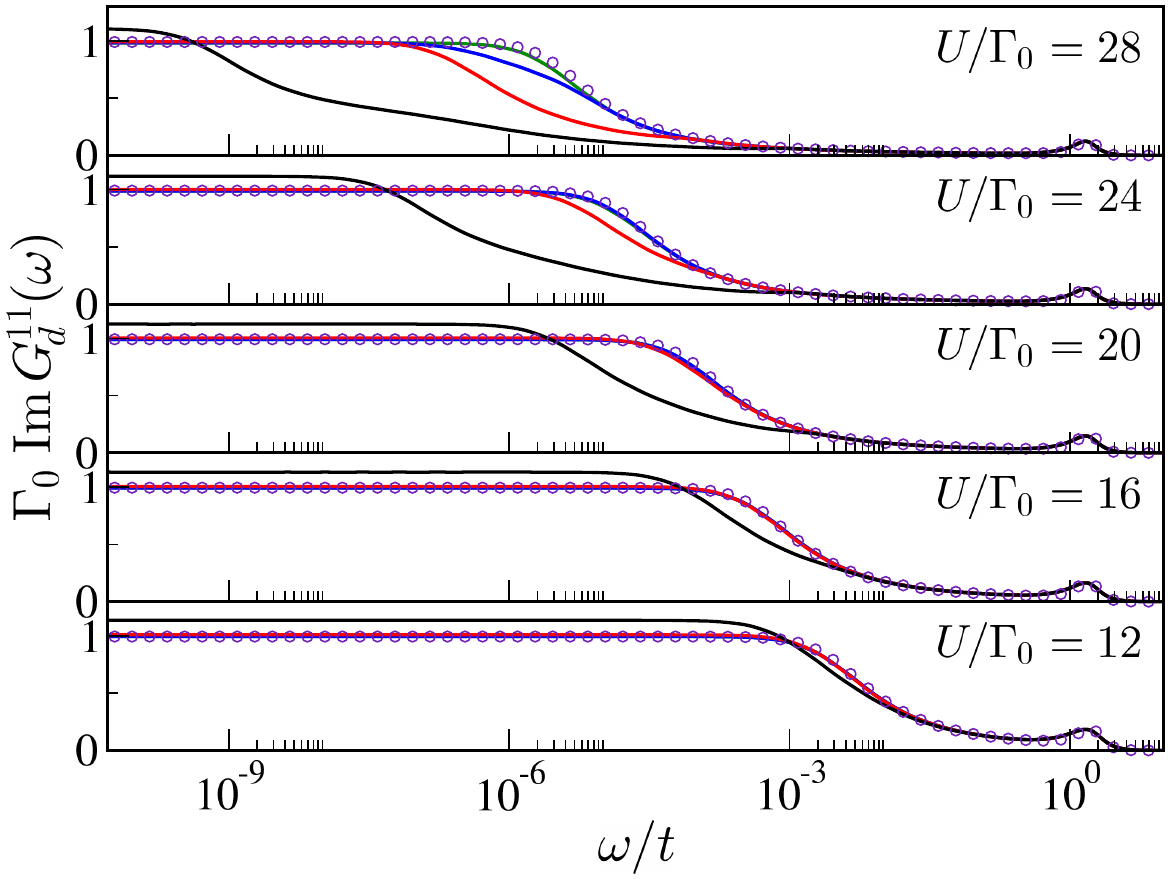}
\caption{\label{fig:hybvar2imp}
Evolution of impurity spectra $-\Gamma_0\text{Im} G_d^{11}(\omega)$ with decreasing $U/\Gamma_0$, obtained by NRG. Plotted at $T=0$ vs $\omega/t$ for $\text{R}_x=\text{R}_y=1,2,3,4 a_{0}$ as the black, red, blue and green lines. Circle points are the corresponding single-impurity results.}
\end{figure}

\sidecaptionvpos{figure}{}
\begin{SCfigure*}
\includegraphics[width=12.1cm]{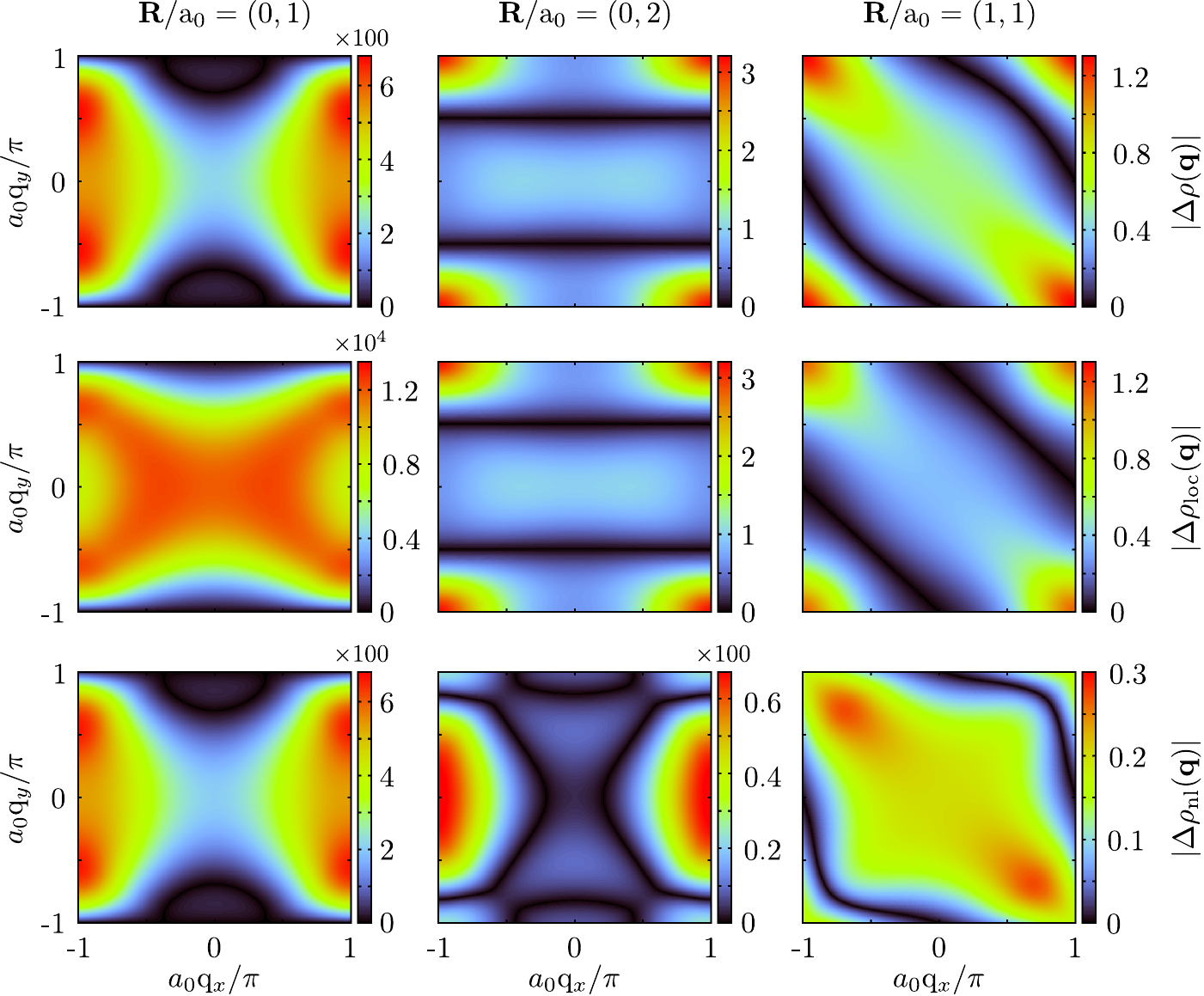}
\caption{\label{fig:2impQPI}
QPI for two equivalent magnetic impurities on the 3d cubic (100) surface, separated by vectors $\textbf{R}/\text{a}_0=(0,1)$, $(0,2)$ and $(1,1)$, as shown in the left, center and right column panels, respectively. The full QPI $|\Delta\rho(\bold q,\omega)|$ is shown in the upper row panels, while the local and nonlocal contributions, obtained from Eqs.~\ref{eq:qpi_q_2imp}, \ref{eq:lam_2imp}, are shown in the center and lower row panels. Obtained for $U/\Gamma_0=28$ as per Fig.~\ref{fig:2imp}, with scanning-energy $\omega=10^{-6}t\equiv 0.1 T_K^{1\text{imp}}$.}
\end{SCfigure*}

\subsection{Effect of impurity hybridization strength} 
\label{sec:2imp_hyb}

As noted in Sec.~\ref{sec:params}, the impurity-host hybridization (and hence single-impurity Kondo temperature) can vary widely in real systems. Here we examine how the physics described in the previous subsection evolves with $U/\Gamma_0$. 

In Fig.~\ref{fig:hybvar2imp}, we consider again the local impurity spectrum $-\Gamma_0\text{Im} G_d^{11}(\omega)$ at $T=0$ with impurity separation $\text{R}_x=\text{R}_y=1,2,3,4a_{0}$, for the range of $U/\Gamma_0$ given in Table~\ref{tab:tk} (keeping $U=3t$ fixed). With increasing hybridization $\Gamma_0$ (and hence increasing $T_K^{1\text{imp}}$), we find that independent Kondo physics becomes relatively more important, overwhelming the effects of inter-impurity interactions at progressively smaller impurity separations (cf Eqs.~\ref{eq:tk}, \ref{eq:rkky}). For $U/\Gamma_0=12$, the impurities behave essentially independently already when separated by $\bold R=(2a_0,2a_0)$.

The interplay between Kondo and RKKY physics is therefore more pronounced for larger $U/\Gamma_0$ where the single-impurity Kondo scale $T_K^{1\text{imp}}$ is smaller.


\subsection{Two-impurity QPI} 
\label{sec:2imp_qpi}

Scattering from two magnetic impurities produces rich structure in the QPI. Two-impurity QPI differs from the idealized single-impurity limit in three key ways. First, the existence of a real-space vector $\bold R=\bold r_2 - \bold r_1$ connecting impurity positions gives rise to additional complex phase factors in the surface scattering t-matrix, see Eq.~\ref{eq:tm_kpar}. Only in the single impurity case can these be eliminated by a suitable basis choice. Second, there are both local ($\alpha = \beta$) and non-local ($\alpha \ne \beta$) contributions to the QPI in Eqs.~\ref{eq:QPI_Q}, \ref{eq:Q_def} due to the matrix structure of the impurity Green functions $\textbf{G}_d(\omega)$. Third, these
Green functions themselves develop rich dynamics due to the interplay between Kondo and effective RKKY physics. As highlighted in Ref.~\onlinecite{pd:qpi1}, the resulting non-trivial scanning-energy and temperature dependence of the QPI is the hallmark of scattering from quantum impurities.

For a generic system containing two equivalent impurities, the QPI can be decomposed into local and non-local contributions, $\Delta\rho(\bold q,\omega)=\Delta\rho_{\text{loc}}(\bold q,\omega) + \Delta\rho_{\text{nl}}(\bold q,\omega)$. Eqs.~\ref{eq:tm_kpar}--\ref{eq:Q_def} imply,
\begin{subequations}
\label{eq:qpi_q_2imp}
\begin{align}
\Delta\rho_{\gamma}(\bold q,\omega) &= -\frac{1}{2\pi\I} [ Q_{\gamma}(\bold q,\omega)-Q_{\gamma}(-\bold q,\omega)^{*} ]\;, \\
Q_{\gamma}(\bold q,\omega) & = V^2 G_d^{\gamma}(\omega) \times \Lambda_{\gamma}(\bold q,\omega) \;,
\end{align}
\end{subequations}
with $\gamma=$`loc' or `nl' [here $G_d^{\text{loc}}(\omega)\equiv G_d^{11}(\omega)=G_d^{22}(\omega)$ and $G_d^{\text{nl}}(\omega)\equiv G_d^{12}(\omega)=G_d^{21}(\omega)$]. The host functions are defined as,
\begin{subequations}
\label{eq:lam_2imp}
\begin{align}
\label{eq:lam_2imp_loc}
\Lambda_{\text{loc}}(\bold q,\omega) &= \cos(\tfrac{1}{2}\bold q\cdot \bold R)\times \Lambda(\bold q,\omega) \;, \\
\label{eq:lam_2imp_nl}
\begin{split}
\Lambda_{\text{nl}}(\bold q,\omega)  &= \int\limits_{1BZ}\frac{d^2\mathrm{\bold{k}}_{\parallel}}{\VBZ}~\cos(\tfrac{1}{2}\bold q\cdot \bold R-\bold k_{\parallel}\cdot \bold R) \\ 
&\qquad  \times G_{0}^{0}(\boldk_{\parallel},\omega)G_{0}^{0}(\boldk_{\parallel}-\bold q,\omega) \;,
\end{split} 
\end{align}
\end{subequations}
where $\Lambda(\bold q,\omega)$ appearing in Eq.~\ref{eq:lam_2imp_loc} is the \emph{single-impurity} result,\cite{pd:qpi1}
\begin{equation}
\label{eq:lam1imp}
\Lambda(\bold q,\omega)  = \int\limits_{1BZ}\frac{d^2\mathrm{\bold{k}}_{\parallel}}{\VBZ}~G_{0}^{0}(\boldk_{\parallel},\omega)G_{0}^{0}(\boldk_{\parallel}-\bold q,\omega) \;.
\end{equation}

Interference between scattering events from each impurity leads to a modulation of the QPI, controlled by the inter-impurity separation vector $\bold R$. In particular, the local contribution $\Delta \rho_{\text{loc}}(\bold q,\omega )$ in Eq.~\ref{eq:lam_2imp_loc} carries an overall factor $\cos(\tfrac{1}{2}\bold q\cdot \bold R)$ which lowers the symmetry of the QPI, and introduces nodal lines of suppressed scattering when $\bold q\cdot \bold R=(2n+1)\pi$. This is demonstrated in the center row panels of Fig.~\ref{fig:2impQPI}, where color plots of $|\Delta \rho_{\text{loc}}(\bold q)|$ are presented for $\bold R/\mathrm{a}_0=(0,1)$, $(0,2)$ and $(1,1)$ in the left, center and right columns, respectively.

By contrast, the relative scattering phase does not appear simply as an overall prefactor to the non-local contribution $\Delta \rho_{\text{nl}}(\bold q,\omega)$ in Eq.~\ref{eq:lam_2imp_nl}. The nodal structure is therefore more complicated, as shown in the lower row panels of Fig.~\ref{fig:2impQPI}.

Importantly, the local and non-local contributions to the total QPI $\Delta \rho(\bold q)$, plotted in the upper row columns, are weighted according to the local and non-local Green function elements --- see Eq.~\ref{eq:qpi_q_2imp}. As discussed in Sec.~\ref{sec:2imp_nrg}, the impurity separation has a major effect on the impurity dynamics. For $\bold R/\mathrm{a}_0=(0,1)$, the non-local contribution to the QPI dominates at low energies $\omega=10^{-6}t$, reflecting the strongly antiferromagnetic RKKY interaction which binds the impurities together in a singlet state. This is shown in the left column panels of Fig.~\ref{fig:2impQPI} (note the rescaled color range); the full QPI $\Delta \rho(\bold q)$ therefore resembles closely its non-local contribution, $\Delta \rho_{\text{nl}}(\bold q)$.

For $\bold R/\mathrm{a}_0=(0,2)$ by contrast, the local contribution dominates (again, note the rescaled color range for the center column panels), and the structure of the full QPI $\Delta \rho(\bold q)$ now resembles closely $\Delta \rho_{\text{loc}}(\bold q)$. As shown in the right column panels, both local and non-local effects are important when the impurities are separated by $\bold R/\mathrm{a}_0=(1,1)$, due to the fine balance of Kondo and RKKY physics in this system.

When the two impurities are well-separated, each is Kondo-screened essentially independently of the other (see Sec.~\ref{sec:2imp_nrg}). This is reflected in the vanishing of non-local impurity Green functions $G_d^{12}(\omega)\simeq 0$, and hence the vanishing of $\Delta \rho_{\text{nl}}(\bold q)$. As shown in Sec.~\ref{sec:2imp_nrg} this happens already for $|\bold R|\gtrsim 4\sqrt{2} \mathrm{a}_0$.


\section{Many Impurities} 
\label{sec:QPImany}

Solutions of full $N$-impurity problems are inherently formidable because they involve $N$ coupled screening channels. In particular, NRG (usually considered the numerical method of choice for quantum impurity problems) cannot traditionally be used for $N> 2$ impurity/channel systems. Recent developments with NRG --- which either utilize model symmetries fully,\cite{awAOP,FeinAuII,awjvd:3bandHund} or exploit the fine-grained RG description obtained by \emph{interleaving} different Wilson chains\cite{akm:mint,akm:mintII} --- significantly reduce the computational cost of solving multichannel problems. Although models with $N=3, 4$ and possibly $5$ channels can now be treated with NRG, true many-impurity systems remain out of reach. Furthermore, such real-space problems cannot generally be cast in the required form involving $N$ impurities connected to the end of $N$ \emph{decoupled} Wilson chains. The matrix Dyson equation, Eq.~\ref{eq:dyson}, can only be diagonalized by a single canonical transformation of operators in special cases where elements of the hybridization matrix possess the symmetry $\Gamma_{ii}(\omega)=\Gamma^{\text{loc}}(\omega)$ and $\Gamma_{i\ne j}(\omega)=\Gamma^{\text{nl}}(\omega)$. Only in this case can one obtain a representation in which $G_{d}^{\nu}(\omega)=(\omega+\I 0^+ -\epsilon_d-\Gamma_{\nu}(\omega)  - \Sigma_{\nu})^{-1}$ are independent, as required for treatment with NRG.

\begin{figure}[t]
\includegraphics[width=7cm]{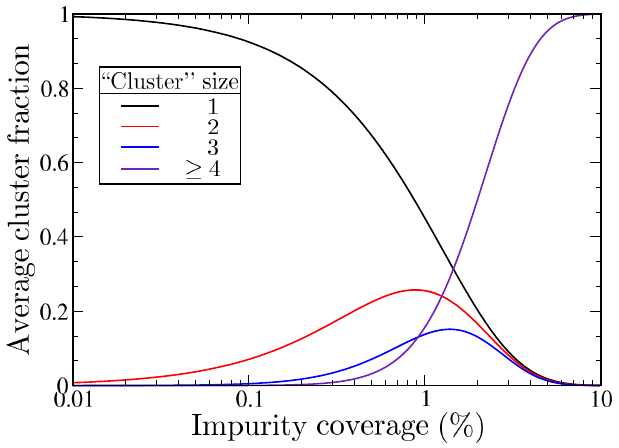}
\caption{\label{fig:impdist}
Average fraction of impurity clusters of size $N_c=1,2,3$ and $\geq 4$, vs impurity coverage on a cubic lattice surface.}
\end{figure}

However, Secs.~\ref{sec:2imp_nrg} and \ref{sec:2imp_hyb} revealed that two impurities behave essentially independently when $|\bold R| > \text{R}_{\text{dil}}$; and that impurities with realistic parameters on the prototypical 3d cubic lattice can be treated independently for a rather small separation $\text{R}_{\text{dil}}\approx 4\sqrt{2}\mathrm{a}_0$. Depending on the specific real-space distribution of impurities, the system could therefore be decomposed into essentially decoupled and hence \emph{independent} smaller clusters. Indeed, if \emph{all} impurities in a given finite sample region are separated by more than $\text{R}_{\text{dil}}$, each impurity can be treated independently. This situation corresponds to the `dilute limit': here an accurate description of the local physics of each impurity is obtained from a model involving a single impurity in the same host. 

For given impurity surface coverage, $\phi = N/N_{s}$ (with $N_{s}$ the number of surface lattice sites), the average inter-impurity separation $\langle \text{R} \rangle \sim a_0 (\phi)^{-1/2}$. But even for low coverage where $\langle \text{R} \rangle \gg \text{R}_{\text{dil}}$, the sample may occasionally contain groups of a few impurities located close enough that their mutual interactions cannot be neglected. These rare `clusters' may nonetheless still be well-separated from other impurities or clusters, whence independent clusters can be treated separately. At a given coverage, and with a specific impurity distribution, one can then describe accurately the many-impurity system in terms of independent impurities and clusters. 

The key question then is: what is the typical occurrence of impurity clusters of size $N_c$ at a given impurity coverage? For given coverage, and surface sample size relevant to FT-STS experiments, the average number of clusters of size $N_c=2,3,4,...$ can then be determined. 

To answer this, Fig.~\ref{fig:impdist} shows the configurationally-averaged fraction of impurity clusters  as a function of surface coverage (with the impurities distributed at random). Specifically, we take a cluster of size $N_c$ to comprise $N_c$ impurities, each within a distance $\text{R}_{\text{dil}}$ of another element of that cluster (but separated from other impurities by more than $\text{R}_{\text{dil}}$). In the following we use $\text{R}_{\text{dil}}=4\sqrt{2}a_0$, although we note from Sec.~\ref{sec:2imp_hyb} that this is a relatively conservative definition of the cluster: larger $T_K^{1\text{imp}}$ leads to smaller $\text{R}_{\text{dil}}$, and hence the earlier onset of the dilute limit. 

At a representative impurity coverage of $0.04\%$, we find that $\sim 96.9\%$ of impurities are independent (i.e. a `cluster' size $N_c=1$), $3\%$ must be treated as an $N_c=2$ impurity pair (c.f. Sec.~\ref{sec:2imp}), and $0.1\%$ are in clusters of $N_c\ge 3$. In a $500 \times 500$ sample region, $0.04\%$ impurity density corresponds to 100 surface impurities: the surface is already rather crowded, with $\langle \text{R} \rangle\sim 28a_{0}$. But even then, there are on average only 3 clusters with $N_c=2$ in such a sample.

We note that impurity coverage can be controlled through various experimental protocols. For example, the extreme dilute limit of $10^{-6}\%$ impurity coverage can be accessed (see e.g. Ref.~\onlinecite{verydil}), and even specific distributions can be realized by manipulation of individual impurities.\cite{CoonCu_Manoharan,passivate} Doped thin films (such as Cr on Cu in Ref.~\onlinecite{CronCufilms}) realize effective 2d systems with impurity densities on the order of $\sim 0.01\%$. Alternatively, evaporation techniques from wires onto surfaces (see e.g. Refs.~\onlinecite{CoonCu_Nikolaus,wahl_dens}) controllably achieve up to $\sim 0.1\%$ impurity density; while dilute alloys\cite{imp_density1,*imp_density2,*imp_density3} are often characterized by an equivalent surface impurity coverage of $0.03-0.3\%$. Higher impurity coverages are more common for certain transition metal surfaces, which can be difficult to clean.

The results of Fig.~\ref{fig:impdist} show that at these experimentally-relevant surface coverages, there is a clear scale separation for cluster occurrence. This suggests a hierarchy of approximation in which the contribution from very rare clusters with size $N_c>N_c^{\text{max}}$ is neglected. Interestingly, at the above typical coverages, many-impurity systems are very well described in terms of independent impurities and independent clusters of size $N_c=2$. Such systems are therefore amenable to treatment with NRG, using only single-impurity results (as in Ref.~\onlinecite{pd:qpi1}) and two-impurity results (as in Sec.~\ref{sec:2imp}).

While the above analysis assumes a random impurity distribution, we add that inter-absorbate interactions in some systems may favor additional clustering. Although the impurity dimer or trimer occurrence for a given average \% coverage might then be enhanced over the prediction of Fig.~\ref{fig:impdist}, the independent cluster approximation nevertheless remains valid.



\begin{figure}[t]
\includegraphics[width=9cm]{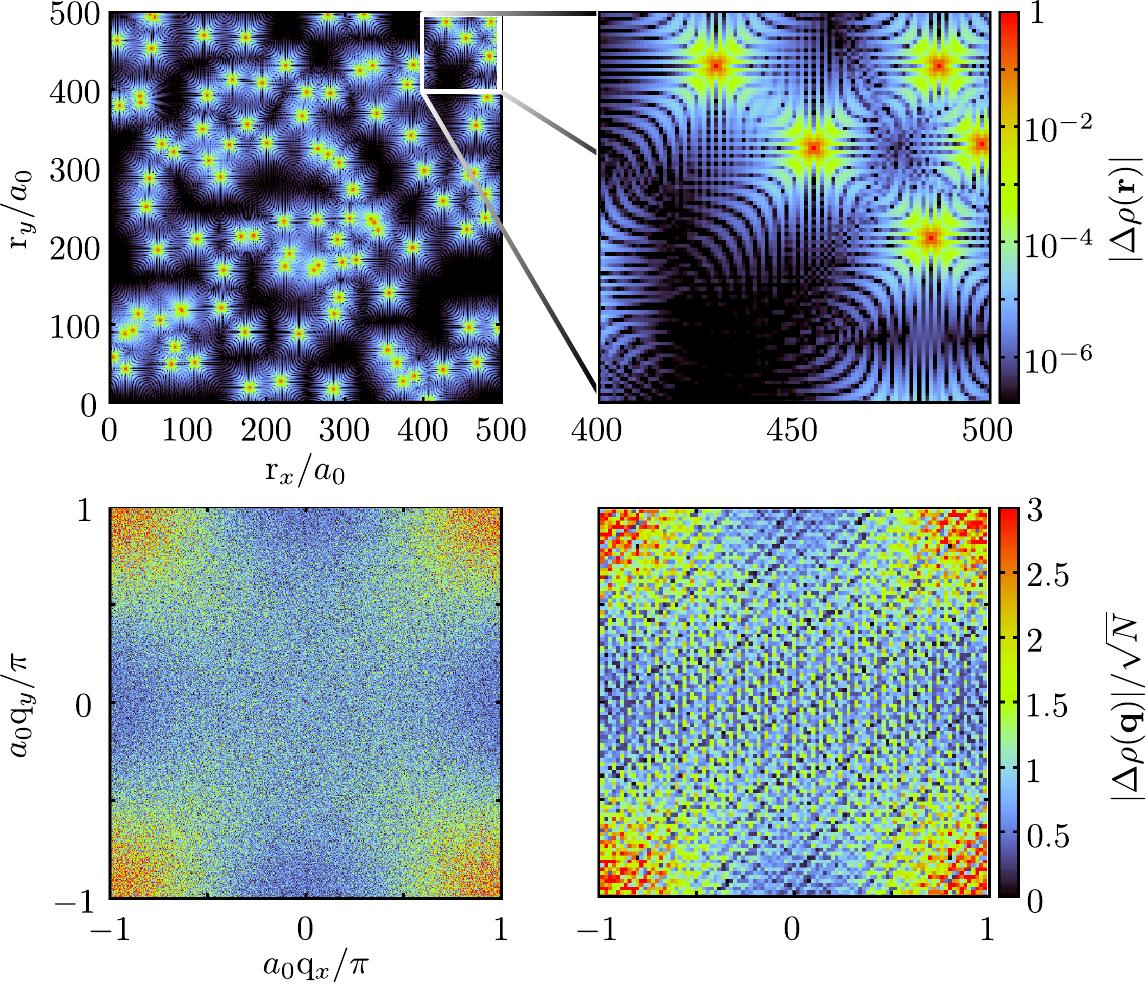}
\caption{\label{fig:manyimp_QPI}
Independent magnetic impurities deposited randomly on a 3d cubic lattice surface. Upper panels: real-space LDOS maps. Lower panels: corresponding QPI. Left panels for $N=100$ impurities within a $500\times 500$ sample region, while right panels show a $100\times 100$ sub-block containing $N=5$ impurities. Calculated for $U/\Gamma_0=28$ and scanning-energy $\omega=10^{-6}t\equiv 0.1T_K^{1\text{imp}}$.
}
\end{figure}


\subsection{Independent impurities: dilute limit} 
\label{sec:indepimp} 

Motivated by the above, we first consider a many-impurity system within an independent impurity picture. Specifically, we study the semi-infinite 3d cubic lattice, with $N=100$ magnetic impurities placed randomly in a $500\times 500$ sample plaquette on the surface (corresponding to $0.04\%$ impurity density).

The independent impurity approximation implies a strictly diagonal hybridization matrix $\boldsymbol \Gamma(\omega) = V^2 G^0(\bold r_i,\bold r_i,\omega)\bold I$. Each impurity then behaves identically to a single quantum impurity in the same host: the impurity distribution does not affect the local impurity physics at this level. Importantly, off-diagonal elements of the t-matrix thus vanish, $T_{\alpha \beta}(\omega)\equiv V^2G_d^{\alpha \beta}(\omega)=\delta_{\alpha\beta} V^2G_d^{\text{loc}}(\omega)$, and the QPI comprises only `local' contributions (c.f. Eqs.~\ref{eq:qpi_q_2imp}, \ref{eq:lam_2imp} for the two-impurity case). 

The impurity distribution does however strongly affect the real-space LDOS map through Eq.~\ref{eq:tmeqn_rs}, even  for independent impurities. The resultant QPI thus also depends on the impurity distribution. It is given generally by 
Eq.~\ref{eq:QPI_Q}, but with (c.f. Eq.~\ref{eq:Q_def}) 
\begin{equation}
\label{eq:indeptimp_qpi}
Q(\boldq,\omega) = T(\boldq,\omega) \times \Lambda(\boldq,\omega) 
\end{equation}
where $\Lambda(\bold q,\omega)$ is the \emph{single-impurity} function Eq.~\ref{eq:lam1imp};
and -- reflecting the independent and identical nature of the impurities --
the full t-matrix $T(\boldq,\omega)$ is the product of a structure factor, 
$S(\boldq)=\sum_\alpha\E^{-\I\boldq\cdot\boldr_\alpha}$, and the local t-matrix 
$ T_\text{loc}(\omega) =V^2G^\text{loc}_d(\omega)$:
\begin{equation}
T(\boldq,\omega)=S(\boldq) \times T_\text{loc}(\omega) 
\end{equation}
The experimentally-measurable power spectrum of the QPI follows as
\begin{equation}
\label{eq:manyimp_cs}
|\Delta\rho(\bold q,\omega)|^2 = \Big ( \sum_{\alpha, \beta} \cos\left [ (\bold r_{\alpha}-\bold r_{\beta})\cdot \bold q \right ]\Big ) \times |\Delta\rho_{1\text{imp}}(\bold q,\omega)|^2 \;,
\end{equation}
where $\Delta\rho_{1\text{imp}}(\bold q,\omega)=-\tfrac{V^2}{\pi}\text{Im}~[G_d^{\text{loc}}(\omega)\times \Lambda(\bold q,\omega)]$ is the pristine single-impurity QPI.\cite{pd:qpi1} As such, the QPI for a many-impurity system in the dilute limit is related simply to the single-impurity QPI, but with an overlaid moir\'{e} pattern due to the superposed cosine factors in Eq.~\ref{eq:manyimp_cs}.

This is demonstrated in Fig.~\ref{fig:manyimp_QPI}, for an $N=100$ impurity system. The upper left panel shows the full spatial map of the LDOS, which allows the impurity positions to be clearly identified. Despite the comparatively high surface density of $0.04\%$ and the apparent crowding, there are only two $N_c=2$ impurity clusters in the $500\times 500$ surface sample (although recall that in this subsection all impurities are treated independently). The upper right panel shows an expanded view of the LDOS for a $100\times 100$ sub-block sample containing $N=5$ impurities. The corresponding QPI for both samples is shown in the lower panels of Fig.~\ref{fig:manyimp_QPI}.

Although the basic structure\cite{pd:qpi1} of the pristine single-impurity QPI survives, the many impurity QPI appears noisy due to the impurity disorder. The $\bold q$-space resolution of the QPI in the lower right panel is naturally more coarse than that in the lower left, because the real-space sample is smaller. The moir\'{e} structure of the QPI due to the $N=5$ impurities is also much more apparent, with well-defined interference fringes characteristic of the underlying impurity separation vectors.


\begin{figure}[t]
\includegraphics[width=8cm]{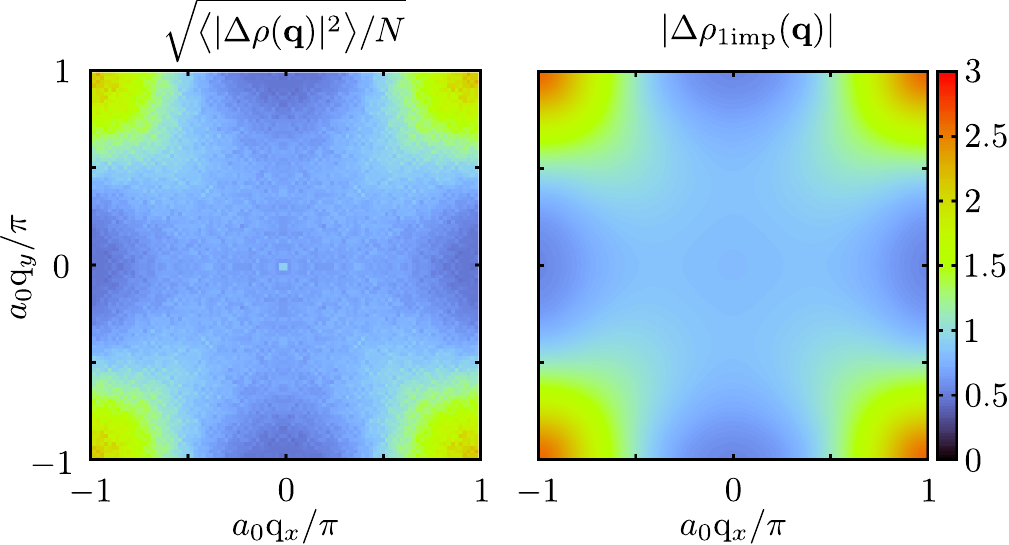}
\caption{\label{fig:blockave}
Left panel: block-averaged and symmetrized QPI, obtained by combining $M=25$ sub-blocks of size $100\times 100$ from the original $500\times 500$ sample shown in the left panels of Fig.~\ref{fig:manyimp_QPI}. The right panel shows the pristine single-impurity result for comparison.
}
\end{figure}
  

For random independent impurities, the QPI calculated via Eq.~\ref{eq:manyimp_cs} from an $L\times L$ surface sample in the limit $L\rightarrow \infty$ recovers the single-impurity result.\cite{03:CapriottiScalapino,06:KodraAtkinson} Equivalently, the single-impurity QPI can in principle be recovered by averaging the QPI over different disorder realizations, since
\begin{equation}
|\Delta\rho_{1\text{imp}}(\bold q, \omega)|^2 =\frac{ \langle|\Delta\rho(\bold q, \omega)|^2\rangle }{ N+N^2\delta_{\bold q,\bold 0} } \;
\end{equation}
where we have used $\langle | S(\bold q) |^2 \rangle=N+N^2\delta_{\bold q,\bold 0}$. Such configurational averaging may not however be experimentally feasible, due to the difficulty in controlling surface topography across disparate real-space regions. 

Instead, one can consider a single $L\times L$ sample region decomposed into $M$ smaller sub-blocks of size $L_b\times L_b$, as in Fig.~\ref{fig:manyimp_QPI}. The QPI from each block can be calculated and the results averaged. If the underlying lattice symmetry is known, the QPI pattern can also be symmetrized by averaging over equivalent scattering vectors $\bold q$. This method reduces  the effect of impurity disorder in the QPI at the expense of $\bold q$-space resolution. This protocol was used to obtain the QPI in the left panel of Fig.~\ref{fig:blockave}, averaging over $M=25$ non-overlapping sub-blocks of $100\times 100$ lattice sites. The result is clearly seen to be quantitatively comparable to the `ideal' single-impurity QPI, shown in the right panel.


\begin{figure}[t]
\includegraphics[width=9cm]{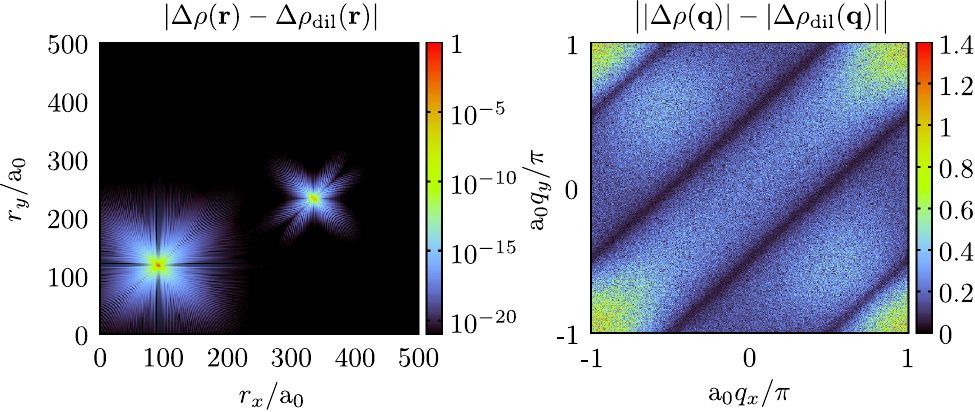}
\caption{\label{fig:qpi_clust}
$N=100$ impurities within a $500\times 500$ surface region of a 3d cubic lattice. Impurity distribution and parameters as in Fig.~\ref{fig:manyimp_QPI}. Left panel: spatially-resolved LDOS difference $|\Delta \rho (\bold r) - \Delta \rho_{\text{dil}}(\bold r)|$, where $\Delta \rho (\bold r)$ is calculated within the independent cluster approximation (treating inter-impurity interactions in the two $N_c=2$ clusters exactly), while $\Delta \rho_{\text{dil}}(\bold r)$ is calculated assuming the dilute limit within an independent impurity picture. Right panel: corresponding QPI difference $|\Delta \rho (\bold q) - \Delta \rho_{\text{dil}}(\bold q)|$.}
\end{figure}
 
\subsection{Independent clusters: beyond the dilute limit} 
\label{sec:indepclust} 

We now go beyond the dilute limit by taking into account explicitly the contribution from impurity clusters whose mutual interactions are important. These clusters are nevertheless well-separated from other impurities or clusters, and therefore remain independent. 

Within this approximation, the hybridization matrix is \emph{block}-diagonal, with $[\boldsymbol \Gamma(\omega)]_{\alpha\ne\beta}$ taken to be non-zero only for impurities $\alpha$ and $\beta$ that are both members of the same cluster. The t-matrix $T_{\alpha\beta}(\omega)$ then has the same block diagonal structure, and the full QPI is as usual computed using Eqs.~\ref{eq:QPI_Q} and \ref{eq:Q_def}.

As a concrete example, we use the same impurity disorder realization as in Fig.~\ref{fig:manyimp_QPI}, corresponding to $N=100$ magnetic impurities within a $500\times 500$ region on the 3d cubic lattice surface (the specific distribution can be visualized through the LDOS modulations in the upper left panel). In this representative case there are two $N_c=2$ clusters, and no larger clusters. The two-impurity clusters are treated independently from the remaining impurities, which are themselves taken to be independent single-impurity problems. Within this framework, the quantum impurity problems are solved using NRG, and the QPI calculated. Since inter-cluster correlations are negligible when clusters are well-separated (Sec.~\ref{sec:2imp_nrg}), this independent cluster approximation is essentially exact.

The local impurity physics of the $N_c=2$ clusters is strongly affected by inter-impurity interactions (Sec.\ \ref{sec:2imp}), compared with isolated or independent impurities. This effect would be observable through spectroscopic local measurements with STM, albeit that the contribution to the overall QPI is expected to be weak because scattering is dominated by the $96\%$ of the impurities in the sample which are effectively independent. 

The left panel of Fig.~\ref{fig:qpi_clust} shows the \emph{difference} between the real-space LDOS calculated using the independent cluster method, and the LDOS obtained through the independent impurity approximation of Sec.~\ref{sec:indepimp}. We find that this difference is highly localized to the cluster sites themselves (note the logarithmic color scale). This confirms that bulk properties are not strongly affected by the contribution from dilute clusters (although an accurate description of \emph{local} quantities does of course require the more sophisticated treatment of intra-cluster interactions).

In the right panel of Fig.~\ref{fig:qpi_clust} we plot the corresponding difference in the QPI, as calculated using the independent cluster and independent impurity approximations. The change in the QPI is dominated by intra-cluster contributions for an $N_c=2$ pair with separation vector $\bold R/\mathrm{a}_0=(2,2)$, as seen directly from the modulation of QPI intensity in the Figure (the impurities in the other $N_c=2$ cluster are separated by $\bold R/\mathrm{a}_0=(4,4)$; intra-cluster interactions are weak in this case, and so its contribution to the overall QPI is small).

Our results indicate that the independent impurity picture, strictly applicable only in the dilute limit, works surprisingly well when the cluster incidence is low. Then the physics of independent quantum impurities dominates the QPI --- even for relatively large impurity densities up to $0.1\%$ when sampling a surface region of size $L\times L$, with $L\sim 100-1000$. To capture inter-impurity interactions, which become important for local quantities near clusters and for fine details in the QPI, the independent cluster method can be employed.


\section{Conclusion} 
\label{sec:conc} 

We have studied systems of multiple magnetic impurities deposited on the surface of a 3d metallic host, and the associated scattering signatures in QPI. From solution of the real-space two-impurity model, we are able to define and identify a `dilute impurity limit' for many-impurity systems, in which each impurity in practice behaves independently of the others. For realistic parameters, the length scale $\text{R}_{\text{dil}}$ for the onset of the dilute limit is found to be strikingly low --- the inter-impurity separation need only be a few lattice sites. Overall, the physics is overwhelmingly dominated by independent single-impurity effects for impurity coverages up to $\sim 0.1\%$ when the Kondo temperature is small, $T_K\sim 0.1K$; while for $T_K\sim 100K$, a surface coverage up to $\sim 1 \%$ is found to remain `dilute'. Measurable QPI therefore reflects the pristine single-impurity result,\cite{pd:qpi1} modulated only by a trivial structure factor due to the real-space impurity distribution. As such, the temperature and scanning-energy dependence of the QPI are entirely characteristic of the underlying single-impurity Kondo effect.

Going beyond the dilute limit to higher impurity coverage, inter-impurity interactions within impurity clusters must naturally be taken into account. The clusters themselves can nevertheless be well-separated, and therefore behave independently. An `independent cluster' approximation, applying in a `dilute cluster limit' then yields accurate results for the physics of such many-impurity systems.

An independent cluster picture must of course ultimately fail when impurity coverage is increased towards and above a percolation threshold where the entire system becomes `connected'. The physics then involves a subtle interplay between Kondo, RKKY, disorder, and the collective heavy fermion physics of the diluted periodic Anderson lattice -- and remains a perenially open problem.


\acknowledgments 
We thank M.~R.~Galpin and R.~Bulla for fruitful discussions. This research was supported by EPSRC grant EP/I032487/1 (AKM,DEL) and the D-ITP consortium, a program of the Netherlands Organisation
for Scientific Research (AKM). We are also grateful to the University of Cologne for the use of HPC facilities.



\begin{thebibliography}{74}
\bibitem [{not()}]{note:clouds}%
  \BibitemOpen
  \bibinfo {note} {Of course, a real-space region containing $\sim 10^{23}$
  conduction electrons can support $\sim 10^{23}$ orthogonal collective states
  capable of screening impurities independently.}\BibitemShut {Stop}%
\end{thebibliography}

%

\end{document}